\documentclass[prl,twocolumn,showpacs,preprintnumbers,amsmath,amssymb]{revtex4}

\usepackage{graphicx}
\usepackage{dcolumn}
\usepackage{bm}
\usepackage{rotating}
\usepackage{color}

\begin{document}

\title{Metastable Feshbach Molecules in High Rotational States}

\author{S. Knoop,$^{1}$ M. Mark,$^{1}$ F. Ferlaino,$^{1,2}$  J. G. Danzl,$^{1}$ T. Kraemer,$^{1}$ H.-C. N\"{a}gerl,$^{1}$ R. Grimm$^{1,3}$}

\affiliation{$^1$Institut f\"{u}r Experimentalphysik and
Forschungszentrum f\"{u}r Quantenphysik, Universit\"{a}t
Innsbruck, 6020 Innsbruck, Austria \\
$^2$ LENS and Dipartimento di Fisica, Universit\`{a} di Firenze, Firenze, Italy\\
$^3$Institut f\"{u}r Quantenoptik und Quanteninformation,
\"{O}sterreichische Akademie der Wissenschaften, 6020 Innsbruck,
Austria}

\date{\today}

\begin{abstract}
We experimentally demonstrate Cs$_2$ Feshbach molecules well above
the dissociation threshold, which are stable against spontaneous
decay on the timescale of one second. An optically trapped sample of
ultracold dimers is prepared in a high rotational state and
magnetically tuned into a region with negative binding energy. The
metastable character of these molecules arises from the large
centrifugal barrier in combination with negligible coupling to
states with low rotational angular momentum. A sharp onset of
dissociation with increasing magnetic field is mediated by a
crossing with a lower rotational dimer state and facilitates
dissociation on demand with a well defined energy.
\end{abstract}


\maketitle

Metastability is at the heart of many phenomena in physics. Energy
barriers or conservation laws can efficiently prevent a system
from decaying into lower-lying states. In the field of ultracold
quantum gases, metastability is ubiquitous. Bose-Einstein
condensates in dilute alkali vapors are intrinsically metastable,
as the absolute ground state of the system is just a tiny piece of
metal. Further prominent examples for the important role of
metastability in ultracold gases can be found in Bose-Einstein
condensates with attractive interactions
\cite{Bradley1997bec,Roberts2001cco}, vortices in rotating
superfluids \cite{Madison2000vfi,Zwierlein2005vas}, dark solitons
\cite{Burger1999dsi,Denschlag2000gsb}, and repulsively bound atom
pairs in optical lattices \cite{Winkler2006rba}.

The association of Feshbach molecules in ultracold quantum gases
\cite{Kohler2006poc} has opened up many new opportunities in the
field, e.g.\ for experiments on quantum states with strong pair
correlations. It has become experimental routine to produce dimers
in a binding energy range of typically 100\,kHz up to a few ten
MHz below the atomic threshold. Considerable efforts are in
progress to extend the energy range of ultracold molecules
initially prepared by Feshbach association to larger binding
energies \cite{Winkler2007cot,Lang2007ctm}, ultimately with the
goal to produce quantum-degenerate molecular samples in the
rovibrational ground state.

\begin{figure}
\includegraphics[width=7cm]{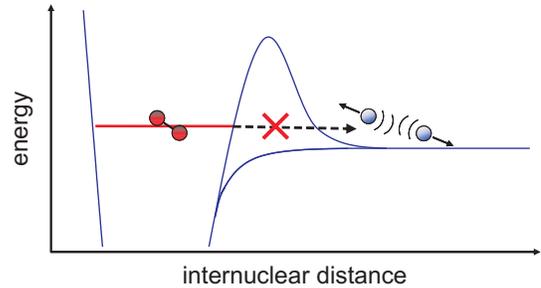} \caption{(Color
online) Illustration of the basic idea of a long-lived metastable
Feshbach molecule in a high rotational state. The solid lines
illustrate two molecular potential curves, with high rotational
angular momentum and with zero rotation. Direct dissociation
through the large centrifugal barrier is strongly suppressed.
Indirect dissociation by coupling at short distances to a partial
wave with low angular momentum is negligible because of the large
difference in the rotational quantum
numbers.}\label{centrifugalbarrier}
\end{figure}

In this Letter, we explore ultracold molecules in a new energy
regime. We create metastable dimers {\em above} the threshold for
spontaneous dissociation into free atoms, i.e.\ dimers with {\em
negative} binding energies. In many experiments, the fast
dissociation of Feshbach molecules above threshold on the time scale
of microseconds \cite{Mukaiyama2004dad, Durr2004dou} is used for
detection purposes by imaging the resulting atoms. Feshbach
molecules with energetically open channels for spontaneous
dissociation were investigated in two previous experiments.
Lifetimes of up to a few tens of milliseconds were demonstrated in
an $s$-wave halo state of $^{85}$Rb$_2$ \cite{Thompson2005sdo} and,
more recently, $^{40}$K$_2$ $p$-wave molecules above threshold
behind a centrifugal barrier were observed with lifetimes of a few
hundreds of microseconds \cite{Gaebler2007pwf}. In contrast to these
rather limited lifetimes, we demonstrate above-threshold molecules
that are stable against dissociative decay on the time scale of a
second; this can be considered infinitely long for experimental
purposes. The basic idea to achieve such metastability is the
transfer of the dimers to a high rotational state where a large
centrifugal barrier in combination with a weak coupling to low
partial-wave scattering states strongly suppresses spontaneous
decay; see illustration in Fig.~\ref{centrifugalbarrier}. Our
experiments are performed on ultracold, trapped Cs$_2$ dimers
transferred into a state with rotational quantum number $\ell=8$,
i.e.\ into an $l$-wave state \cite{Russell1929ron}; here the
centrifugal energy barrier is as high as $h \times 150$\,MHz.

\begin{figure}
\includegraphics[width=7.5cm]{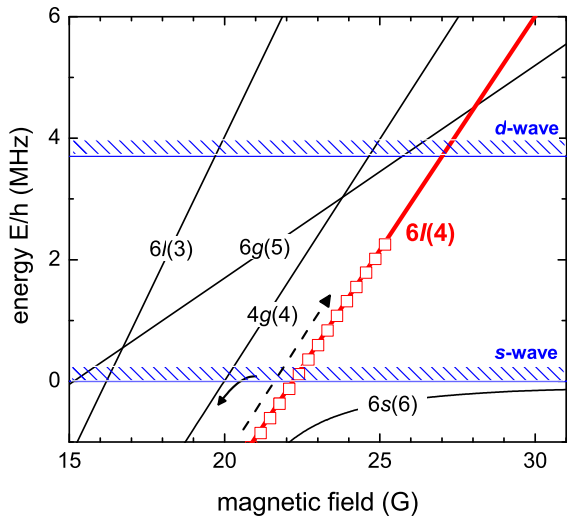} \caption{(Color online)
Energy curves of the relevant Cs$_2$ dimer states as function of
the magnetic field \cite{Chin2004pfs,Mark2007sou}. The $s$-wave
threshold ($E=0$) and the height of the $d$-wave centrifugal
barrier ($E=h\times3.7$\,MHz) are indicated. The quantum state
relevant for this work is labeled with $6l(4)$. The open squares
represent spectroscopic data on this state obtained from measuring
its magnetic moment, and the arrows indicate the path for
preparation of the dimers (for details see
text).}\label{energydiagram}
\end{figure}

The two-body scattering properties of Cs and the underlying
molecular structure have been thoroughly investigated in previous
work \cite{Chin2004pfs,Mark2007sou}. Feshbach resonances up to
$g$-wave character have been found. The underlying coupling of the
$s$-wave scattering continuum to $g$-wave molecular states is
special to cesium because of relatively strong indirect spin-spin
coupling \cite{Leo2000cpo}. This second-order interaction in
general allows coupling between different partial-wave states with
a difference $\Delta \ell$ in rotational quantum numbers up to
$|\Delta \ell| = 4$. The Cs dimer energy structure in the relevant
range is shown in Fig.~\ref{energydiagram}. The $s$-wave threshold
corresponds to two free atoms in the lowest hyperfine ground state
sublevel $|F,m_F\rangle=|3,3\rangle$ with zero kinetic energy. The
notation $f\ell(m_f)$ for molecular states was introduced in
Ref.~\cite{Mark2007sou}; the symbols $f$ and $m_f$ represent the
quantum numbers for the total internal angular momentum and its
projection.

The Cs$_2$ spectrum provides several accessible $l$-wave states,
which have been identified in
Refs.~\cite{Chin2005oof,Mark2007siw,Mark2007sou}. Remarkably,
these states do not manifest themselves in observable Feshbach
resonances in the scattering of atoms, which is a consequence of
the negligible coupling to the scattering continuum for $|\Delta
\ell| > 4$. The $l$-wave states can nevertheless be populated
efficiently by using coherent state-transfer at avoided level
crossings \cite{Mark2007sou}, where each step obeys the selection
rule $\Delta \ell \le 4$ for the rotational angular momentum.

In this work, the state $6l(4)$ serves as a model system for
metastable dimers in high rotational states above the dissociation
threshold. The starting point for our experiments is a sample of
typically $1\times10^4$ $l$-wave dimers prepared in a crossed-beam
CO$_2$ laser trap at a temperature of $T$=250(50)\,nK; the mean trap
frequency is $\bar{\omega}=2\pi\times 41(1)$\,Hz. The experimental
procedures have been developed earlier \cite{Mark2007sou}. In brief,
molecules are first created by Feshbach association in the state
$4g(4)$ using the 19.84-G resonance, followed by transfer into
$6l(4)$ via an intermediate state $6g(6)$; the latter state has a
binding energy of about $h\times5$\,MHz and is out of the energy
range displayed in Fig.~\ref{energydiagram}. The state $6l(4)$ has a
magnetic moment of $0.96(1)\mu_B$ and crosses the dissociation
threshold at a magnetic field of $22.0(2)$\,G. We have
spectroscopically investigated this state by extending
magnetic-moment measurements \cite{Mark2007sou} into energy regions
above threshold; see open squares in Fig.~\ref{energydiagram}.

To measure the lifetime of the $l$-wave state, we hold the sample at
magnetic fields corresponding to energies above and below the
dissociation threshold. After variable storage times of up to
0.4\,s, the remaining molecules are quickly (within typically
10\,ms) subjected to the reverse transfer route and dissociated at
the $g$-wave Feshbach resonance at 19.84\,G. Standard absorption
imaging is finally applied to the resulting atom sample
\cite{Herbig2003poa}.

\begin{figure}
\includegraphics[width=8.5cm]{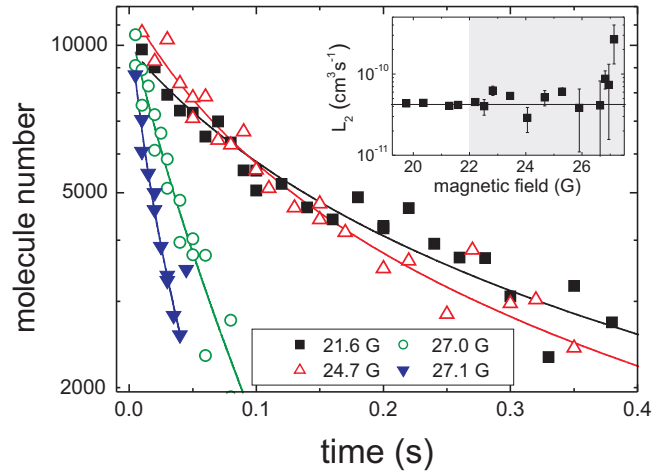} \caption{(Color online)
Loss measurements of the $6l(4)$ molecular sample for different
magnetic fields. One set of data is taken below the dissociation
threshold ($B = 21.6$\,G); three further sets ($24.7$\,G, $27.0$\,G,
$27.1$\,G) refer to situations above threshold. The lines are fits
to the measurements including two-body and one-body decay. The inset
shows the resulting two-body loss coefficient $L_2$, where the
shaded region indicates the magnetic field region above threshold.
The horizontal line indicates the average $L_2$ below
threshold.}\label{decaycurves}
\end{figure}

In Fig.~\ref{decaycurves}, we show typical decay measurements.
Just below threshold ($B=21.6$\,G, filled squares), we observe a
non-exponential decay of the molecular sample on a timescale of
100\,ms, which is a result of inelastic two-body collisions. For
an energy of $h\times2$\,MHz above threshold (24.7\,G, open
triangles), the behavior is very similar with a slightly faster
decay. Only at magnetic fields around 27\,G we observe a much
faster loss of molecules. We analyze the decay measurements using
the rate equation
\begin{math}\dot{N}/{N}=-\alpha-L_2 \bar{n},\end{math}
where $L_2$ is the loss coefficient that describes two-body decay
resulting from inelastic dimer-dimer collisions. The parameter
$\alpha$ represents the rate of spontaneous dissociation as the
dominant one-body decay process of the dimers. The mean molecular
density $\bar{n}$ is related to the molecule number $N$ by
$\bar{n}=(m\bar{\omega}^2/(2\pi k_B T))^{3/2}N$, where $m$ is the
mass of a Cs atom. Below threshold the loss of molecules is solely
determined by inelastic two-body decay and the one-body decay term
can be omitted. Other loss sources, such as light-induced
dissociation and background collisions, can be neglected under our
experimental conditions.

The fit results for $L_2$ as a function of the magnetic field are
shown in the inset of Fig.~\ref{decaycurves}. Below threshold no
magnetic field dependence is observed. Here $L_2$ has a value of
$4.2(0.2)_{\text{stat}}(1.4)_{\text{syst}}\times10^{-11}$\,cm$^3$s$^{-1}$,
similar to the values for Cs$_2$ found in other quantum states
\cite{Chin2005oof, Zahzam2006amc}. The systematic error is based
on the uncertainty in the temperature and the trap frequency
measurements. Above threshold $\alpha$ is a free parameter in the
fit procedure and the competition between one-body and two-body
decay introduces a large error. The values obtained for $L_2$
above threshold scatter around the value obtained below threshold
with a trend to somewhat higher values.

\begin{figure}
\includegraphics[width=8cm]{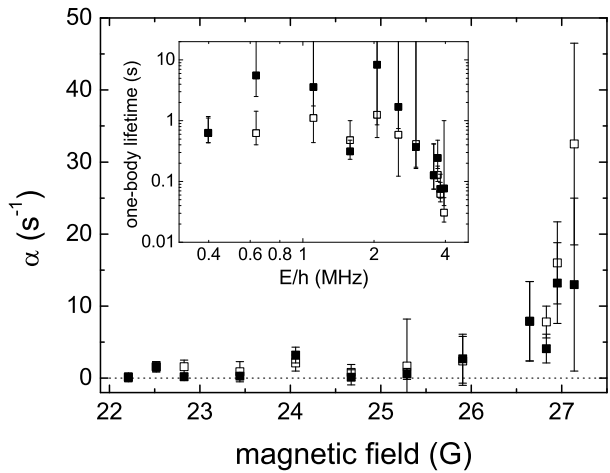} \caption{Dissociation
rate $\alpha$ for the $6l(4)$ state as function of magnetic field
obtained from an unconstrained fit (filled squares) and a
constrained fit, fixing $L_2$ to its value below threshold (open
squares). The inset shows the corresponding one-body lifetime as
function of energy above threshold.}\label{dissociationrate}
\end{figure}

The results for the dissociation rate $\alpha$ are shown in
Fig.~\ref{dissociationrate} (filled symbols). We also apply a
constrained fit, in which $L_2$ is fixed to its constant value below
threshold. The results of this constrained fit (open symbols) give
an upper limit for the dissociation rates under the plausible
assumption that collisional decay does not decrease above threshold.
Below 26\,G, i.e.\ up to 4\,G above threshold, the dissociation
rates remain close to zero. Around 27\,G, however, an increase is
observed, which is most clearly seen from the constrained fit.
One-body lifetimes for spontaneous dissociation in the regime of
negative binding energies above threshold are obtained by inverting
the dissociation rates and are shown in the inset of
Fig.~\ref{dissociationrate} as function of the energy above
threshold. The one-body lifetimes up to 26\,G ($h\times3$\,MHz) are
at least 1\,s, which highlights the long-lived metastable character
of the $l$-wave Feshbach molecules. Beyond this time scale, we
cannot rule out a very slow dissociative decay, e.g.\ by a
high-order coupling to the $s$-wave scattering continuum.

\begin{figure}
\includegraphics[width=8cm]{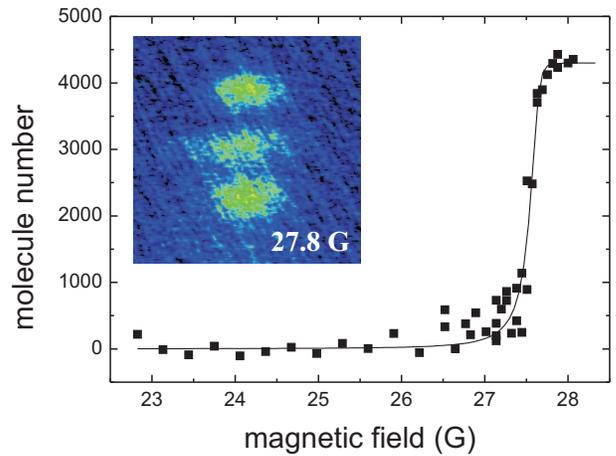} \caption{(Color
online) The number of dissociated molecules as function of the
magnetic field. The curve is a fit, based on modeling the amount of
$g$-wave character around the crossing between the $6l(4)$ state and
the $6g(5)$ state. The inset shows an absorption image of rapid
dissociated $6l(4)$ dimers at a magnetic field of 27.8\,G (averaged
over 16 shots).}\label{ldissociation}
\end{figure}

The increase of the dissociation rate around 27\,G shows that a new
dissociation channel opens up. This behavior cannot be explained by
tunneling through the $l$-wave centrifugal barrier, as the estimated
rate for this process is four orders of magnitude smaller than
observed here experimentally. We explain the onset of dissociation
by coupling to a $g$-wave state, which mediates the dissociation
\cite{shapereso}. The $6l(4)$ state crosses a state $6g(5)$ at about
28\,G (see Fig.~\ref{energydiagram}). When the crossing is
approached, a $g$-wave component starts to mix in. This increasing
admixture then allows decay into $s$- and $d$-waves without the
requirement of any higher-order coupling beyond the indirect
spin-spin interaction.

To investigate the dissociation process around 27\,G in more detail,
we directly image the resulting atoms \cite{Volz2005fso,
Gaebler2007pwf}, without the reverse transfer route and dissociation
at 19.84\,G. In Fig.~\ref{ldissociation}, we show the number of
dissociated molecules as function of the magnetic field. Here the
images are taken 5\,ms after reaching the final magnetic field value
in a fast ramp. Below 26\,G no dissociated molecules are observed,
while in the region of 27-28\,G a rapid appearance is seen. The
inset of Fig.~\ref{ldissociation} shows an absorption image at
$27.8$\,G. The dissociation pattern is consistent with an
interference between $s$- and $d$-waves. The presence of $d$-waves
is expected as the dissociation occurs above the $d$-wave
centrifugal barrier.

We use a simple two-channel avoided crossing approach to model the
magnetic field dependence of the dissociation rate. Near the
crossing, we describe the molecular state as a superposition of the
$l$- and $g$-wave states. The $g$-wave amplitude determines the
dissociation rate and increases smoothly from zero to one when the
avoided crossing of quasi-bound states is followed from below. A fit
based on this model is shown by the solid line in
Fig.~\ref{ldissociation}. The model allows us to determine the
position of the crossing within an uncertainty of 0.1\,G. We obtain
a value of 27.7\,G, which is consistent with our previous knowledge
of the molecular spectrum \cite{Mark2007sou}.

The dissociation mechanism discussed for the $6l(4)$ state also
agrees with observations in further experiments on the two other
neighboring $l$-wave states $6l(3)$ and $6l(5)$. The latter state
crosses the atomic threshold at 35.0(2)\,G and has a first avoided
crossing with a $g$-wave state only around $h\times12$\,MHz (not
shown in Fig.~\ref{energydiagram}). Here we explored magnetic fields
up to 48\,G and corresponding energies up to $h \times 8$\,MHz
without finding any significant decay, even far above the $d$-wave
centrifugal barrier \cite{otherlwave}. In contrast, the $6l(3)$
state shows significant dissociation already at a relatively small
energy of $h\times0.7$\,MHz above threshold \cite{Mark2007siw}. This
is explained by the presence of a crossing with the $6g(5)$ state,
which is found at a magnetic field only 0.5\,G higher than the
intersection of the $6l(3)$ state with the $s$-wave threshold (see
Fig.~\ref{energydiagram}).

To conclude, we have demonstrated the metastable character of
ultracold dimers in a high rotational state above the dissociation
threshold, where the binding energy is negative. In our
experiments, we observed Cs$_2$ Feshbach molecules in an $l$-wave
state to be stable against dissociative decay on a time scale of
at least one second. The large centrifugal barrier suppresses
tunneling to the $l$-wave scattering continuum while the coupling
to lower partial waves is extremely small. The metastable region
and onset of dissociation for the $l$-wave states depends on the
location of the first crossing with a $g$-wave state, which
mediates dissociative decay.

In future work, shielding the dimers against collisional decay can
be achieved by trapping in an optical lattice
\cite{Thalhammer2006llf}. In such a periodic environment one may
also create a Mott-like state with exactly one molecule per
lattice site \cite{Volz2006pqs}. This opens up the possibility to
create novel metastable quantum states with strong pair
correlations. The possibility to achieve dissociative decay on
demand with a well-defined energy by switching the magnetic field
to particular values opens up further possibilities, e.g.\ for a
controlled collective decay of the inverted medium in analogy to
superradiance.

We thank S. D\"{u}rr and T. Volz for fruitful discussions. We
acknowledge support by the Austrian Science Fund (FWF) within SFB
15 (project part 16). S.~K.\ is supported within the Marie Curie
Intra-European Program of the European Commission. F.~F.\ is
supported within the Lise Meitner program of the FWF.


\begin{thebibliography}{27}
\bibitem[{6l5()}]{otherlwave}
\bibinfo{note}{The $6l(5)$ state shows even more favorable conditions regarding
metastability compared to the $6l(4)$ state. However, we focus on
the $6l(4)$ state in this paper, as the production and detection
of $6l(5)$ state is more difficult.}
\end{thebibliography}
\end{document}